\documentclass[prd,nofootinbib,preprintnumbers]{revtex4}
\usepackage[utf8]{inputenc}
\usepackage{amsfonts,amsmath,amssymb}
\usepackage{graphicx}
\usepackage{color}
\usepackage[plainpages=false, colorlinks=true, anchorcolor=blue, linkcolor=blue, citecolor=blue, bookmarks=false]{hyperref}
\usepackage{natbib}
\usepackage{enumitem}
\usepackage{subcaption}
\newcommand{\rthis}[1]{\textcolor{black}{#1}}
\begin{document}
\newcommand{\apjl}{Astrophys. J. Lett.}
\newcommand{\apjs}{Astrophys. J. Suppl. Ser.}
\newcommand{\aap}{Astron. \& Astrophys.}
\newcommand{\aj}{Astron. J.}
\newcommand{\araa}{Ann. Rev. Astron. Astrophys. } 
\newcommand{\mnras}{Mon. Not. R. Astron. Soc.}
\newcommand{\apss} {Astrophys. and Space Science}
\newcommand{\jcap}{JCAP}
\newcommand{\pasj}{PASJ}
\newcommand{\pasa}{Pub. Astro. Soc. Aust.}
\newcommand{\physrep}{Physics Reports}
\title{An independent search for annual modulation and its significance in ANAIS-112 data}
\author{Aditi \surname{Krishak}$^1$}\altaffiliation{E-mail:aditi16@iiserb.ac.in}

\author{Shantanu  \surname{Desai}$^2$ }  
\altaffiliation{E-mail: shntn05@gmail.com}

\affiliation{$^{1}$ Department of Physics, Indian Institute of Science Education and Research, Bhopal, Madhya Pradesh 462066, India}

\affiliation{$^{2}$Department of Physics, Indian Institute of Technology, Hyderabad, Telangana-502285, India}

\begin{abstract}
We perform an independent search for  sinusoidal-based  modulation in the recently released ANAIS-112 data, which could be induced by dark matter scatterings. We then evaluate  this hypothesis against the null hypothesis that the data contains only background, using four different model comparison techniques. These include frequentist, Bayesian, and two information theory-based criteria (AIC and BIC). This analysis was done on both the residual data (by subtracting the exponential fit obtained from the ANAIS-112 Collaboration) as well as the total (non-background subtracted) data.  We find that according to the  Bayesian model comparison test, the null hypothesis of no modulation is decisively favored over    a cosine-based annual modulation for the non-background subtracted dataset in 2-6 keV energy range. None of  the other model comparison tests decisively favor any one hypothesis over another. This is the first  application of Bayesian and information theory techniques to test the annual modulation hypothesis in ANAIS-112 data, extending our previous work on the DAMA/LIBRA and COSINE-100 data.  Our analysis codes have also been made publicly available.
\end{abstract}

\maketitle

\section{Introduction}
The ANAIS collaboration~\cite{ANAIS}  (A19, hereafter) recently released their first scientific results, related to  testing the long-standing DAMA claim of annual modulation caused by dark matter scatterings~\cite{DAMA18} (and references therein).
Their target material consists of  112.5 kg of NaI (hence the experiment has been named ANAIS-112) and the total lifetime of the data released in 2019 was 1.5 years. With the current data, the ANAIS-112 data were found to be  consistent with the null hypothesis of no modulation, with $p-$values of 0.65 and 0.16 in the 2-6 and 1-6 keV energy intervals respectively.

In two recent works~\cite{Krishak,Krishak2}, we performed an independent search for annual modulation, using data from  two similar direct dark matter detection experiments, namely DAMA~\cite{DAMA18} and COSINE-100~\cite{Cosine}. In these works, we evaluated  the significance of the annual modulation using four independent model comparison techniques: frequentist, information theoretical and Bayesian analysis. We now carry out the same exercise on the recently released ANAIS-112 data (which has been kindly made available to us by the collaboration). Our analysis is therefore complementary to the model comparison tests carried out in A19.

The manuscript is organized as follows.
A brief summary of the ANAIS-112 results can be found in Sect.~\ref{sec:ANAIS}. Our analysis and results of the same data is described in Sect.~\ref{sec:analysis}.  A comparison to our previous results using DAMA and COSINE-100 data can be found in Sect.~\ref{sec:comparison}. We conclude in Sect.~\ref{sec:conclusions}. We have made our analysis codes publicly available and they can be found at \url{https://github.com/aditikrishak/ANAIS112_analysis}

\section{ANAIS-112 results}
\label{sec:ANAIS}
We now recap the ANAIS-112 results from A19, wherein more details can be found. 
The ANAIS-112 experiment consisting of 112.5 kg of NaI as target, is located at the Canfranc Underground Laboratory, LSC, in Spain under 800~m of rock overburden. The experiment uses nine NaI modules. The experiment started taking data in August 2017 and has released about 1.5 years of data until February 2019.
The background rates in the modules were fit to a superposition of  constant and exponential terms.
The annual modulation search was done in two different energy bins, viz. [1-6] and [2-6] keV. 
These fits were done by binning the data in 10-day intervals. The function used for fitting both the potential signal and background is given by (we use the same notation as in A19):
\begin{equation}
R(t) = R_0 + R_1 \exp(-t/\tau)  + A \cos \omega (t+\phi),     
\label{eq:1}
\end{equation}
\noindent where $R_0$, $R_1$, and $\tau$ are used to parameterize the exponential background; and $A$, $\omega$, and $\phi$ correspond to the amplitude, angular frequency and phase of the expected signal respectively.  While doing the fits, the period and the phase were fixed to 1 year and -62.2 days respectively, where the phase corresponds to the expected maximum around June 2nd. For doing a background only fit, $A$ was assumed to be equal to zero, and for fitting the data to the signal, $A$ is assumed to be a free parameter. An independent search for signal with the  phase as a free parameter was also done using this data.

The data was found to be  consistent with the null hypothesis in both the  2-6 keV  and 1-6 keV energy intervals, corresponding to $p$-values of 0.67 and 0.18 respectively. The corresponding $p$-values  for the annual modulation hypothesis are 0.65 and 0.16 respectively. The best fits were inconsistent with the DAMA/LIBRA best fits at 2.5$\sigma$ and 1.9$\sigma$ in the two energy intervals. More details of these results are available in A19. We note that  ANAIS, with the present data, has not yet reached the sensitivity to test the annual modulation effect reported by the DAMA/LIBRA experiment. However, we would like to do a proof of principles demonstration of Bayesian and information theory based techniques to test the hypothesis of annual modulation in the ANAIS-112 data.

\section{Our analysis}
\label{sec:analysis}
We first do an annual modulation fit by using the same background parameters as those used by  the ANAIS collaboration, and carry out hypothesis testing using only the residual rates. We then vary the background parameters, and do a combined fit to both the signal and background.  Once parameter estimation is done for both the hypotheses, we carry out model  comparison against the null hypothesis of no modulation. We describe parameter estimation in Sect.~\ref{sec:pm} and present results of model comparison in Sect.~\ref{sec:mc}.

\subsection{Parameter estimation}
\label{sec:pm}
  We denote the cosine modulation as hypothesis $H_1$ and the background only hypothesis as $H_0$.
 For both 1-6 keV and 2-6 keV intervals, we first do  a fit to the residual ($y(t)$) obtained by subtracting the constant and exponential background, using the best-fit background parameters from  A19.
For $H_1$, the residual $y(t)$ is modeled as:
 \begin{equation}
    y(t) = A\cos\omega(t+\phi), 
    \label{eq:yt}
\end{equation}
where $A$, $\omega$, and $\phi$ have the same meaning as in Eq.~\ref{eq:1}. For the null hypothesis $H_0$, $y(t)$ is fit to a constant value. The residual $y(t)$ has been plotted in Fig.~2 of A19 for both the 1-6  keV and 2-6 keV energy intervals.

We also do a   fit directly to the total event rate by using the same equation used in A19 (cf. Eq.~\ref{eq:1}), by finding the best-fit parameters for both the signal and background. This fit was done from the total event rates in both the 1-6 keV and 2-6 keV energy ranges.

\begin{figure*}
\begin{subfigure}{.75\textwidth}
  \centering
  \includegraphics[width=\linewidth]{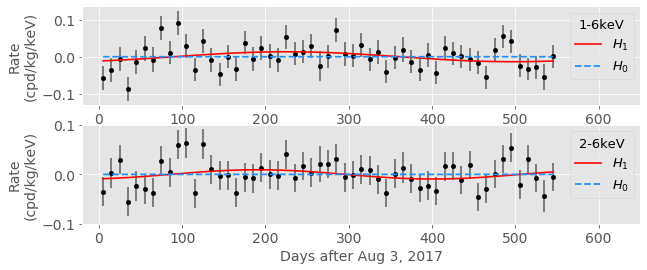}
  \caption{}
  \label{fig:fig1a}
\end{subfigure}
\begin{subfigure}{.75\textwidth}
  \centering
  \includegraphics[width = \linewidth]{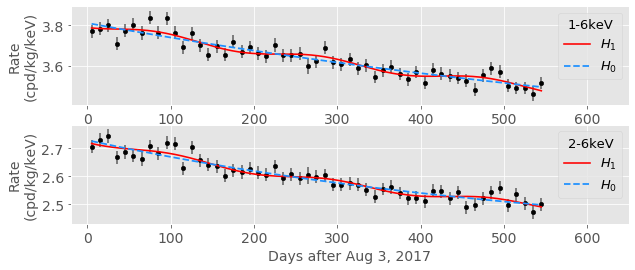}
  \caption{}
  \label{fig:fig1b}
\end{subfigure}
\caption{Plots showing (a) 
best fit for the data with exponential background subtracted, $y(t)$, and (b) best fit for $R(t)$. The $H_0$ corresponds to the background hypothesis of no modulation, whereas the $H_1$ hypothesis corresponds to a cosine modulation. The data points along with error bars have been obtained courtesy the ANAIS collaboration.}
\label{fig1}
\end{figure*}

To find the best-fit parameters for both the datasets, we construct a  $\chi^2$ function, which quantifies the differences between the model and the data for both the residual and total event rates as follows:
\begin{eqnarray}
\chi^2 &=& \sum_{i=1}^N\left( \frac{d_i-R(t)}{\sigma_{di}}\right)^2 , \\
\label{eq:chisqtot}
\chi^2 &=& \sum_{i=1}^N\left( \frac{f_i-y(t)}{\sigma_{fi}}\right)^2 , \\
\label{eq:chisqres}
\end{eqnarray}
where $N$ is the total number of data points; $d_i$ and $f_i$ denote the data vector for the total and residual event rate respectively; $\sigma_{di}$ and $\sigma_{fi}$ encapsulate the total error in $d_i$ and $f_i$ respectively, which have been provided to us by the ANAIS-112 collaboration. The free parameters which we vary in Eq.~\ref{eq:chisqtot} for $R(t)$  to minimize $\chi^2$ are: $R_0$, $R_1$, $\tau$, $A$, $\omega$, $\phi$. The corresponding free parameters  for $y(t)$ are $A$, $\omega$, and $\phi$.
The null hypothesis  corresponds to a constant value for $y(t)$ when we fit the residuals, and to $A=0$ in Eq.~\ref{eq:1}, when we fit for $R(t)$.

For $H_0$ and $H_1$ in both the energy intervals, we obtain the best-fit parameters by $\chi^2$ minimization using least-square optimization methods in the \texttt{scipy Python} module, and then carry out model comparison using multiple methods. While doing the minimization, the period is constrained between the width of the time bin (10 days) and the maximum duration of the dataset (545 days). Our fits are not sensitive to values outside this range.
The parameter $\tau$ in  $R(t)$ has also been given a positive constraint.
The best-fit values we get for both the hypotheses along with their 1$\sigma$ errors can be found in Table~\ref{table:1} and ~\ref{table:1a} for the residuals and total rate respectively. In Table~\ref{table:1}, we also show the best-fit value for the amplitude obtained in A19 from the residual data. We note that our best-fit amplitude differs from that in A19 by about 1.3$\sigma$. This is because we have varied all the three parameters in the cosine function, whereas in A19  the period and phase were fixed to the DAMA best-fit values and only the amplitude was allowed to be free. Because of the degeneracy between the parameters, there is some difference between our best-fit amplitude and those found in A19.
The plots showing the residual data  as well as the total data  (along with the best fits for both $H_0$ and $H_1$) are shown in Figs.~\ref{fig:fig1a} and Figs.~\ref{fig:fig1b} respectively.

\subsection{Model Comparison}
\label{sec:mc}
When multiple functions can explain the data, we are faced with assessing the relative viability of each of  the models. There is no unique way to carry out such an assessment
and one can come up with different criterion to compare two models. There are two broad schools of thought on this issue~\cite{Sanjib,Liddle,Liddle07}. The Bayesian analysis compares the probability of the model given the data, whereas the frequentist method compares the expected predictive accuracy of the two models for future data. The information theory technique are a distinct class and have both frequentist and Bayesian interpretations~\cite{Liddle07}.
More details comparing these techniques and the pitfalls of each of them
can be found in Refs.~\cite{Liddle,Liddle07,Shi,Trotta,astroml,Sanjib,Weller}. Among these plethora of techniques, Liddle recommends the use of Bayesian methods, since it updates the prior probability to the posterior probability and has an unambiguous interpretation therein~\cite{Liddle07}.  In Cosmology, nowadays only Bayesian methods are used for model comparison, whereas in experimental particle physics,  frequentist methods usually get used. For direct dark matter search experiments only a handful of works have used Bayesian and information theoretical techniques to assess the statistical significance of annual modulation~\cite{Krishak,Krishak2,Messina}.

In this work, we apply all the three types of techniques to test the robustness of our results.
The results from each of these sets of tests are outlined below. For brevity, we skip the theory behind the  model comparison tests, which can be found in our earlier companion works~\citep{Krishak,Krishak2} and references therein.

\begin{table}
    \centering
    
    \begin{tabular}{|c|c|c|}
    
    \hline
    {}&{\textbf{1-6 keV}} & {\textbf{2-6 keV}}\\
        \hline 
         \multicolumn{3}{|c|}{\textbf{Modulation only}}\\
        \hline

        $A$ (cpd/kg/keV) & -0.0135 $\pm$ 0.0066 & -0.0094 $\pm$ 0.0056 \\
        $\omega$ (radians/day) & 0.0115 $\pm$ 0.0033 & 0.0149 $\pm$ 0.0032 \\
        $\phi$ (days)& 47 $\pm$ 93 & 20 $\pm$ 76\\
               \hline 
               \multicolumn{3}{|c|}{\textbf{ANAIS-112 Best fit}}\\ \hline
        $A$ (cpd/kg/keV) & -0.0015 $\pm$ 0.0063 & -0.0044 $\pm$ 0.0058 \\  \hline

 \end{tabular}
\caption{The best-fit parameter values obtained for the cosine modulation hypothesis for  the background-subtracted residual data ($y(t)$), when fitted using Eq.~\ref{eq:yt}. The background-subtracted residual data was obtained from the ANAIS-112 collaboration. The last row indicates the best-fit value for the amplitude obtained in A19, wherein the period (0.01721 rad/day) and phase (-62.2 days) have been kept fixed to the DAMA best-fit values.}
\label{table:1}
\end{table}

\begin{table}
    \centering
    
    \begin{tabular}{|c|c|c|}
    
    \hline
    {}&{\textbf{1-6 keV}} & {\textbf{2-6 keV}}\\
        \hline 
         \multicolumn{3}{|c|}{\textbf{Exponential background + modulation}}\\
        \hline
        \textbf{$R_0$} (cpd/kg/keV)&2.74 $\pm$ 1.15 &2.34 $\pm$ 0.10 \\
        \textbf{$R_1$} (cpd/kg/keV)& 1.05 $\pm$ 1.14&4.00 $\pm$ 0.10  \\
        \textbf{$\tau$} (days) &1541 $\pm$ 2002 &591 $\pm$ 251    \\
        \textbf{$A$} (cpd/kg/keV)& 0.0186 $\pm$ 0.0065 & 0.0107 $\pm$ 0.0042 \\
        \textbf{$\omega$} (radians/day) & 0.0322 $\pm$ 0.0025 &  0.0330 $\pm$ 0.0028\\
        \textbf{$\phi$} (days) & -84 $\pm$ 18 & -101 $\pm$ 20\\
             \hline 
 \end{tabular}
\caption{The best-fit parameter values obtained for the 
cosine modulation hypothesis 
from $\chi^2$ minimization for the total event rate ($R(t)$) from Eq.~\ref{eq:1}.  The total event rate data was obtained from the ANAIS-112 collaboration.}
\label{table:1a}
\end{table}

\begin{itemize}
\item {\bf Frequentist model comparison:}
 In this test (also known as the likelihood ratio test~\cite{Weller}), while comparing two models, the model with the larger value of $\chi^2$ pdf would be considered as the favored model among the two~\cite{Ganguly}. The $\chi^2$ pdf for a given value of $\chi^2$ and degrees of freedom equal to $k$ is given by:
 \begin{equation}
P(\chi^2|k) = \frac{1}{2^{k/2}\Gamma (k/2)}\left(\chi^2\right)^{k/2-1} \exp (-\chi^2/2)      \label{eq:chisq} 
 \end{equation}
 The $\chi^2$ pdf (or likelihood) for  both $H_0$ and $H_1$ for the residual rate can be found in Table~\ref{table:2}. We note that in the 1-6 keV range, the $\chi^2$ likelihood for the $H_0$ hypothesis is marginally greater than than the $H_1$, whereas the opposite is true for the 2-6 keV  energy interval.  
Making use of the fact that the two models are nested, we  quantify the $p$-value of the cosine model as compared to the background model using Wilk's theorem~\cite{Wilks}.  We note that Wilk's theorem assumes that the  data is asymptotic and all the additional free parameters in $H_1$ are not on the boundary~\cite{Lyons}. Additional caveats and limits of applicability of this test can be found in Ref.~\cite{Kashyap}.
For our example, the difference in $\chi^2$ between the two models satisfies a $\chi^2$ distribution with degrees of freedom equal to three for the total event rate and two for the the residual rate.  The $p$-value can be evaluated from the $\chi^2$ c.d.f. as discussed in Ref.~\cite{Ganguly}. The corresponding significance or $Z$-score is calculated by finding the number of standard deviations by which  a Gaussian variable would fluctuate in one direction to give the corresponding $p$-value \cite{Cowan,Ganguly}. 

The $\chi^2$ values per degree of freedom and the  likelihood of the model, given by the $\chi^2$ pdf can be found in Table~\ref{table:2}, along with the $p$-value and $Z$-score. As we can see, the $H_1$ (background + cosine modulation) is very marginally favored, with a significance of only  $1.39\sigma$ for 1-6 keV and $0.59\sigma$ for 2-6 keV. The corresponding results when the background parameters are allowed to vary can be found in Table~\ref{table:3}. For the 1-6 keV  interval, $H_1$ is very  marginally favored over $H_0$, whereas the converse is true in the  2-6 keV interval.  However, the difference in significance in both the hypotheses is marginal ($1.1-1.8\sigma$). 

\item{\bf AIC and BIC:}
The Akaike and Bayesian information criterion are two information theory-based criterion used for model comparison~\cite{Liddle07}, where additional terms get added to  $\chi^2$ to penalize for the additional free parameters.
\begin{eqnarray}
AIC &=& \chi^2+ 2p \\
BIC &=& \chi^2+ p\ln N
\label{eq:AICBIC}
\end{eqnarray}
where $p$ is the total number of free parameters and $N$ is the total number of data points.

While comparing two models, the one with the smaller value of AIC and BIC is preferred. The significance can be evaluated using the qualitative strength of evidence rules given in Ref.~\cite{Shi}. We   note that one assumption in applying BIC is that the posterior pdf is Gaussian.

The $\Delta$AIC and $\Delta$BIC values are tabulated in Table~\ref{table:2} and Table~\ref{table:3} for residual and total rates respectively. For residual rates, $\Delta$BIC gives ``positive'' evidence in favor of $H_0$ in both the energy bands; while $\Delta$AIC gives ``substantial'' support for $H_0$ in 1-6 keV and for $H_1$ in 2-6 keV range. When we fit for the total rates~(cf. Table~\ref{table:3}), we infer that the  $\Delta$BIC values in the 1-6 keV and 2-6 keV region point to ``positive'' and ``strong'' evidence respectively,  for  the  null hypothesis of no modulation.  For the same data $\Delta$AIC  points to ``substantial''  evidence, in support of cosine modulation in 2-6 keV, and in support of the null hypothesis in 1-6 keV using the same strength of evidence rules. However, for none of the datasets, the absolute difference in AIC and BIC between the two models  crosses the threshold of 10 (needed for any one model to be decisively favored). So the AIC and BIC tests do not decisively prefer any one model.

\item \textbf {Bayesian Model Comparison:}
We carry out a Bayesian model comparison by calculating the Bayesian odds ratio, which in this case is equal to the  Bayes factor  $B_{21}$ for the $M_2$ model in comparison to the $M_1$ hypothesis. We note that unlike previous methods, this model comparison technique does not use the best-fit values of the parameters.
Here, we consider the null hypothesis ($H_0$) to be $M_1$ and the cosine model ($H_1$) to be $M_2$. $B_{21}$ is given by
\begin{equation}
B_{21} = \frac{P(M_2 | D)}{P(M_1|D)} ,
\end{equation}
 where $P(M_2 | D)$ and $P(M_1|D)$ are the posterior probabilities for $M_2$ and $M_1$ respectively given the observed data $D$. The Bayesian evidences for both $H_0$ and $H_1$ have been evaluated using the {\tt Dynesty} package~\cite{Speagle} in {\tt Python}. Therefore, if the cosine hypothesis is favored, the Bayes factor is greater than one and vice-versa.
 
  To calculate the evidence for both the datasets, we assume a Gaussian likelihood given by:
  \begin{equation}
     P(D|M)=\prod_{i=1}^N \frac{1}{\sigma_i\sqrt{2\pi}} \exp -\left[\frac{(\eta_i-f(x,\theta))^2}{2\sigma_i^2}\right],
  \end{equation}
  where  $\eta_i$ is the observed data; $f(x,\theta)$ is the model function used to fit the data; and $\sigma_i$ is the observed error in $\eta_i$. This likelihood  assumes that the error residuals are Gaussian.  In order to calculate  the Bayesian evidence for the cosine signal, we use three different priors. First we use uniform priors for period and phase. We then also calculate the Bayesian evidence  by choosing a Gaussian prior on the period with mean and standard error equal to the DAMA best-fit value of about a year and standard  deviation determined by the error in DAMA best-fit period equal to ($0.999 \pm 1$ year)~\cite{DAMA18}. Finally, we also calculate the Bayesian evidence by choosing a Gaussian prior on the phase with mean and standard deviation equal to DAMA's best-fit values for the phase $(145 \pm 5)$ days~\cite{DAMA18}. These three sets of priors are assumed for model comparison with both the residual and the total rate. For the null hypothesis, we use uniform priors for all the background parameters. A tabular summary of the priors used for both residual and total event rates can be found in Tables~\ref{tab:priortable1} and Tables~\ref{tab:priortable2} respectively.

 The resulting Bayes factor can be found in Tables~\ref{table:2} and~\ref{table:3} for the residual and total event rates respectively. For the background subtracted data,  (cf. Table~\ref{table:2}) the Bayes factor is slightly greater than one for both the energy ranges using all three prior choices, indicating that the modulation hypothesis is favored. However, the absolute value of the Bayes factor   is less than five, which implies that according to Jeffreys' scale~\cite{Trotta}, the difference between the two models is negligible. For the non-background subtracted data (cf. Table~\ref{table:3}), the Bayesian evidence for the null hypothesis is much greater than that for the cosine modulation. In the 1-6 keV region, the values of the Bayes factor is $\mathcal{O} (10^{-2})$ for all the three priors,  and from the Jeffreys scale~\cite{Trotta} it shows very strong to decisive evidence for the null hypothesis.\footnote{Note that in Table ~\ref{table:3}, we have tabulated the ratio of Bayesian evidence for modulation hypothesis to null hypothesis, so if the null hypothesis is decisively favored, the Bayes factor should be less than 0.01}. For the 2-6 keV energy range, the values of the Bayes factor for all three priors is $\mathcal{O} (10^{-3})$, and therefore the null hypothesis is decisively favored.
 
 To summarize our results of Bayesian model comparison, we find that for the total event rate data, the Bayes factor decisively favors the null hypothesis in the 2-6 keV range and also strongly/decisively in the 1-6 keV range, whereas with the residual rate data, the difference between the two models is negligible. One possible reason for the difference in results between the two datasets  is that the difference in free parameters between the two hypotheses is equal to three, whereas for the residual rate it is two. Bayes factor harshly penalizes models with extra free parameters~\cite{Trotta,Sanjib}. Another possible reason is that since the residual rate data is distributed around zero (cf. Fig.~\ref{fig:fig1a}) with both positive and negative excursions, it is comparatively easier to fit a sinusoidal function with all its parameters free, as opposed to a flat background.

\end{itemize}

\begin{table*}
\begin{tabular}{|cccc|}
\hline \hline
\textbf{Prior} &   $\mathbf{A}$ & \boldmath$\omega$ & \boldmath$\phi$
\\
&  (cpd/kg/keV) & (radian/day) & (days) \\ \hline  
P1  & $\mathcal{U}$ (-max($|f_i|$),max($|f_i|$))  & $\mathcal{U}$ (0.0115,0.6168)  &$\mathcal{U}$ (0,365) \\ 
P2&   $\mathcal{U}$ (-max($|f_i|$),max($|f_i|$))  & $\mathcal{U}$ (0.0115,0.6168)  &$\mathcal{N}$ (145,5) \\ 
P3 & $\mathcal{U}$ (-max($|f_i|$),max($|f_i|$))  & $\mathcal{N}$ (0.0172,$1.36 \times 10^{-5}$)  &$\mathcal{U}$ (0,365) \\ 
\hline
\end{tabular}
\caption{Priors used for the calculation of Bayesian evidence from the residual rates  ($y(t)$) for the different terms in  Eq.~\ref{eq:yt} for the cosine modulation hypothesis. For the null hypothesis the last two terms are not used, and the prior for the constant term is same as that for $A$.  The uniform prior on $\omega$ corresponds to period between 10 and 545 days.}
\label{tab:priortable1}
\end{table*}

\begin{table*}
\begin{tabular}{|ccccccc|}
\hline \hline
\textbf{Prior} & $\mathbf{R_0}$ & $\mathbf{R_1}$ & \boldmath$\tau$ & $\mathbf{A}$ & \boldmath$\omega$ & \boldmath$\phi$
\\
& (cpd/kg/keV) & (cpd/kg/keV) & (days) & (cpd/kg/keV) & (radian/day) & (days) \\ \hline  
P1 & $\mathcal{U}$ (0,max($d_i$)) &$\mathcal{U}$ (0,max($d_i$)) &$\mathcal{U}$ (0.1,550) & $\mathcal{U}$ (0,max($d_i$))  & $\mathcal{U}$ (0.0115,0.6168)  &$\mathcal{U}$ (0,365) \\ 
P2& $\mathcal{U}$ (0,max($d_i$)) &$\mathcal{U}$  (0,max($d_i$)) &$\mathcal{U}$ (0.1,550) & $\mathcal{U}$ (0,max($d_i$))  & $\mathcal{U}$ (0.0115,0.6168)  &$\mathcal{N}$ (145,5) \\ 
P3 & $\mathcal{U}$ (0,max($d_i$)) &$\mathcal{U}$ (0,max($d_i$)) &$\mathcal{U}$ (0.1,550) & $\mathcal{U}$ (0,max($d_i$))  & $\mathcal{N}$ (0.0172,$1.36 \times 10^{-5}$)  &$\mathcal{U}$ (0,365) \\ 
\hline
\end{tabular}
\caption{Priors used for the calculation of Bayesian evidence when considering the total event rates ($R(t)$), for the different terms in  Eq.~\ref{eq:1} for the oscillation hypothesis. For the null hypothesis the last three terms are not used, whereas the same priors are used for the first three. }
\label{tab:priortable2}
\end{table*}

\begin{table}
    \centering
    
    \begin{tabular}{|c|cc|cc|}
    
    \hline
    {}&\multicolumn{2}{c|}{\textbf{1-6 keV}}&\multicolumn{2}{c|}{\textbf{2-6 keV}}\\
        {} & {$H_0$} & {$H_1$}& {$H_0$} & {$H_1$}\\
        \hline
        \textbf{Frequentist} & {} & {} &{}&{}\\
         {$\chi^2$}/DOF & 65.9/54 & 61.0/52 & 49.4/54 & 46.8/52\\
         {$\chi^2$ pdf } &  0.0176 & 0.0235 & 0.0377 & 0.0376\\
         {$p$-value}    & \multicolumn{2}{c|}{0.08} & \multicolumn{2}{c|}{0.28}\\
         {significance} & \multicolumn{2}{c|}{1.39 $\sigma$}& \multicolumn{2}{c|}{0.59 $\sigma$} \\
         \hline
         \textbf{AIC} & 67.9 & 66.9 & 51.4 & 52.8\\
         {$\Delta$ AIC} & \multicolumn{2}{c|}{-1.0}& \multicolumn{2}{c|}{1.4} \\
         \hline
         \textbf{BIC}  & 69.9 & 72.9 & 53.4 &58.9\\
         {$\Delta$ BIC}  & \multicolumn{2}{c|}{3.0}&\multicolumn{2}{c|}{5.5}\\
         \hline
         \textbf{Bayes Factor} & \multicolumn{4}{c|}{$P(H_1|D)/P(H_0|D)$} \\
         P1 & \multicolumn{2}{c|}{4.5 }&\multicolumn{2}{c|}{3.2}\\
         P2 & \multicolumn{2}{c|}{4.5 }&\multicolumn{2}{c|}{3.0}\\
         P3 & \multicolumn{2}{c|}{2.2 }&\multicolumn{2}{c|}{3.0}\\
         \hline 
 \end{tabular}
\caption{Summary of model comparison results for the residual rates ($y(t)$) using frequentist, Bayesian, and information theoretic criterion for $H_0$ (background only) and $H_1$ (cosine modulation).  The Bayes factor is the ratio of Bayesian evidence for $H_1$ hypothesis to $H_0$ hypothesis, so a values greater than one will prefer the $H_1$ hypothesis. The Bayes factor has been calculated using three different sets of priors: P1, P2, and P3 (cf. Tab.~\ref{tab:priortable1}). No one hypothesis is decisively  favored using any of the model comparison tests.}
\label{table:2}
\end{table}

\begin{table}
    \centering
    
    \begin{tabular}{|c|cc|cc|}
    
    \hline
    {}&\multicolumn{2}{c|}{\textbf{1-6 keV}}&\multicolumn{2}{c|}{\textbf{2-6 keV}}\\
        {} & {$H_0$} & {$H_1$}& {$H_0$} & {$H_1$}\\
        \hline
        \textbf{Frequentist} & {} & {} &{}&{}\\
         {$\chi^2$}/DOF & 59.5/52 & 50.7/49 & 47.9/52 & 42.4/49\\
         {$\chi^2$ pdf} &  0.0267 & 0.0383 & 0.0389 & 0.0363\\
         {$p$-value}    & \multicolumn{2}{c|}{0.03} & \multicolumn{2}{c|}{0.13}\\
         {significance} & \multicolumn{2}{c|}{1.8$\sigma$}& \multicolumn{2}{c|}{1.1$\sigma$} \\
         \hline
         \textbf{AIC} & 65.5 & 62.7 & 53.9 & 54.4\\
         {$\Delta$ AIC} & \multicolumn{2}{c|}{-2.8}& \multicolumn{2}{c|}{0.5} \\
         \hline
         \textbf{BIC}  & 71.5 & 74.7 &59.9&66.4\\
         {$\Delta$ BIC}  & \multicolumn{2}{c|}{3.2}&\multicolumn{2}{c|}{6.5}\\
         \hline
         \textbf{Bayes Factor} & \multicolumn{4}{c|}{$P(H_1|D)/P(H_0|D)$} \\
         P1 & \multicolumn{2}{c|}{0.013} &\multicolumn{2}{c|}{ 0.004}\\
         P2 & \multicolumn{2}{c|}{0.008} &\multicolumn{2}{c|}{ $8 \times 10^{-4}$}\\
         P3 & \multicolumn{2}{c|}{0.015} &\multicolumn{2}{c|}{ 0.001}\\
         \hline 
 \end{tabular}
\caption{Summary of model comparison results using frequentist, Bayesian and information theoretic criterion for $H_0$ (background only) and $H_1$ (background+cosine modulation). Here, the fit is done to the total count rate, including background and signal.  The Bayes factor  (for 2-6 keV energy range) decisively favor the null hypothesis for all the three prior choices. In the 1-6 keV, interval also, the Bayes factor strongly/decisively favors the null hypothesis depending on the prior choice. The other tests do not decisively favor any one hypothesis.}
\label{table:3}
\end{table}

\section{Comparison with Cosine-100 and DAMA/LIBRA}
\label{sec:comparison}
We have done a similar model comparison analysis of annual modulation for DAMA/LIBRA~\cite{Krishak} as well as COSINE-100~\cite{Krishak2}, using all the four metrics discussed here. The DAMA/LIBRA experiment has accumulated over 1 ton-year worth of exposure~\cite{DAMA18}. For DAMA/LIBRA we find that the statistical significance using all four tests provides decisive evidence for annual  modulation. The $Z$-score from frequentist test exceeds 5$\sigma$; both $\Delta$AIC and $\Delta$BIC exceed 10; and the Bayes factor exceeds the value of 100~\cite{Krishak}.  Therefore, all the four tests point to a concordant picture.

The COSINE-100 experiment has at the time of their data release about the same exposure as the ANAIS-112 experiment, of 97.7 kg years~\cite{Cosine}. For the COSINE-100 data, we did two sets of model selection tests. We first varied all the parameters in the cosine function. We then did another fit with the period kept fixed at DAMA's best-fit value. In both the cases we used the total COSINE-100 event rate, where we did a fit to both the background as well as the cosine modulation.
For the both cases, we find that the frequentist and AIC tests  \rthis{are only marginally different between the two hypotheses with frequentist significance of 0.42$\sigma$/0.14$\sigma$ and $\Delta $AIC of 2.4/2.6 depending on whether $\omega$ is a free parameter or fixed~\cite{Krishak2}}.
However, the BIC and Bayesian model comparison decisively/strongly favor the null hypothesis of no modulation \rthis{with $\Delta$BIC equal to 12.5 (8.9) and natural logarithm of the Bayes factor equal to -16.0 (-7.0)} depending 
on whether the angular frequency as kept as a free parameter or frozen to the best-fit value found by the DAMA collaboration~\cite{Krishak2}. This is somewhat similar to what we see for ANAIS-112, where when we analyze the total count rate data, the Bayes factor   shows decisive/strong evidence in favor of the null hypothesis. Similarly for ANAIS-112 also, the  frequentist and AIC test do not conspicuously differentiate between the two models. \rthis{A summary table comparing all these metrics for DAMA/LIBRA, COSINE-100 and ANAIS-112 for data in the 2-6 keV energy range for the same assumptions can be found in Table~\ref{tab:comparison}.}

\begin{table}
\centering
\begin{tabular}{|c|c|c|c|}
\hline 
& \textbf{DAMA/LIBRA} & \textbf{COSINE-100} & \textbf{ANAIS-112} \\ \hline
Freq. $p$-value & $6 \times 10^{-31}$  & 0.34  & 0.13  \\ 
Freq. significance &11.5$\sigma$ & 0.4$\sigma$  & 1.1$\sigma$    \\ \hline 
$\Delta$AIC & -117 & 2.6 & 0.5 \\ \hline 
$\Delta$BIC & -112 & 12.5  & 6.5 \\ \hline 
Bayes factor  & $2.8 \times 10^{17}$ & $1.1\times 10^{-7}$ & 0.004   \\
\hline
\end{tabular}
\caption{\rthis{A comparison of  the three model comparison metrics for DAMA/LIBRA, COSINE-100 and Anais-112 between background and cosine modulation for data in the 2-6 keV range using the same assumptions. The definitions of $\Delta$AIC, $\Delta$BIC, and Bayes factor is same as in Table~\ref{table:2}.
More details on the DAMA/LIBRA and COSINE-100 results can be found in our companion works~\cite{Krishak,Krishak2}. The ANAIS-112 values are replicated from Table~\ref{table:3}
corresponding to 2-6 keV.}}
\label{tab:comparison}
\end{table}

\section{Conclusions}
\label{sec:conclusions}

The ANAIS-112 dark matter direct detection experiment consisting of 112.5 kg of NaI, which has been designed to test the long-standing DAMA annual modulation claim, recently released their first results using 1.5 years of data, having a total exposure of 157.55 kg year~\cite{ANAIS}. This data  was found by the collaboration to be consistent with  the null  hypothesis of no annual modulation.

As a follow-up to  our previous work with DAMA and COSINE-100 data~\cite{Krishak,Krishak2}, we carried out an independent search for  annual modulation using the same data in two different energy intervals: 1-6  and 2-6 keV. For each of these energy intervals, we fitted both the total event rate and also the background-subtracted residual event rate for a cosine modulation, which could be induced by dark matter interactions. In the latter case, we used the background subtracted data, provided by the ANAIS collaboration.

We then carried out  a model comparison analysis of these different data sets to test if the current ANAIS-112 data is compatible with annual  modulation. For this purpose, we used four different model comparison techniques: frequentist, Bayesian, AIC, and BIC.  For Bayesian model comparison we used three sets of priors, listed in Tables~\ref{tab:priortable1} and ~\ref{tab:priortable2}. A tabular summary of our results can be found in Tables~\ref{table:2} and~\ref{table:3}.

When we analyze the background subtracted data, no one hypothesis is decisively favored using any of the model comparison tests used herein.
When we analyze the total event rate, the AIC and frequentist model comparison test cannot robustly discriminate between the two hypotheses. However, using the same data  BIC provides strong evidence in the 2-6 keV energy interval for no modulation over a cosine-based  modulation.  The Bayesian model comparison test however decisively favors the null hypothesis of no modulation in the  2-6 keV energy intervals for all the three priors. In the 1-6  keV range, the Bayesian test provides strong/decisive evidence for the null hypothesis depending on the choice of the prior.
Therefore, using the current data, only the Bayesian evidence test  decisively favors the null hypothesis of no modulation in the 2-6 keV, when we analyze the total event rate data. 

This is a proof of principles application of Bayesian and information theory based techniques to the ANAIS-112 data (extending our previous work on the DAMA/LIBRA and COSINE-100 data) and is complementary to the model comparison techniques carried out by the ANAIS-112 collaboration. It is straightforward to  apply these techniques to the same dataset with increasing exposure. To improve transparency in data analysis, we have made publicly available our analysis codes and they can be found at \url{https://github.com/aditikrishak/ANAIS112_analysis}.

\begin{acknowledgements}
Aditi Krishak is supported by a DST-INSPIRE fellowship. We are grateful to the ANAIS Collaboration for providing  us the raw data from A19 for our analysis.
\end{acknowledgements}

\bibliography{main}

\begin{thebibliography}{19}
\expandafter\ifx\csname natexlab\endcsname\relax\def\natexlab#1{#1}\fi
\expandafter\ifx\csname bibnamefont\endcsname\relax
  \def\bibnamefont#1{#1}\fi
\expandafter\ifx\csname bibfnamefont\endcsname\relax
  \def\bibfnamefont#1{#1}\fi
\expandafter\ifx\csname citenamefont\endcsname\relax
  \def\citenamefont#1{#1}\fi
\expandafter\ifx\csname url\endcsname\relax
  \def\url#1{\texttt{#1}}\fi
\expandafter\ifx\csname urlprefix\endcsname\relax\def\urlprefix{URL }\fi
\providecommand{\bibinfo}[2]{#2}
\providecommand{\eprint}[2][]{\url{#2}}

\bibitem[{\citenamefont{Amaré et~al.}(2019)}]{ANAIS}
\bibinfo{author}{\bibfnamefont{J.}~\bibnamefont{Amaré}} \bibnamefont{et~al.},
  \bibinfo{journal}{Phys. Rev. Lett.} \textbf{\bibinfo{volume}{123}},
  \bibinfo{pages}{031301} (\bibinfo{year}{2019}), \eprint{1903.03973}.

\bibitem[{\citenamefont{{Bernabei} et~al.}(2018)\citenamefont{{Bernabei},
  {Belli}, {Bussolotti}, {Cappella}, {Caracciolo}, {Cerulli}, {Dai},
  {d'Angelo}, {Di Marco}, {He} et~al.}}]{DAMA18}
\bibinfo{author}{\bibfnamefont{R.}~\bibnamefont{{Bernabei}}},
  \bibinfo{author}{\bibfnamefont{P.}~\bibnamefont{{Belli}}},
  \bibinfo{author}{\bibfnamefont{A.}~\bibnamefont{{Bussolotti}}},
  \bibinfo{author}{\bibfnamefont{F.}~\bibnamefont{{Cappella}}},
  \bibinfo{author}{\bibfnamefont{V.}~\bibnamefont{{Caracciolo}}},
  \bibinfo{author}{\bibfnamefont{R.}~\bibnamefont{{Cerulli}}},
  \bibinfo{author}{\bibfnamefont{C.-J.} \bibnamefont{{Dai}}},
  \bibinfo{author}{\bibfnamefont{A.}~\bibnamefont{{d'Angelo}}},
  \bibinfo{author}{\bibfnamefont{A.}~\bibnamefont{{Di Marco}}},
  \bibinfo{author}{\bibfnamefont{H.-L.} \bibnamefont{{He}}},
  \bibnamefont{et~al.}, \bibinfo{journal}{Nuclear Physics and Atomic Energy}
  \textbf{\bibinfo{volume}{19}}, \bibinfo{pages}{307} (\bibinfo{year}{2018}),
  \eprint{1805.10486}.

\bibitem[{\citenamefont{{Krishak} et~al.}(2020)\citenamefont{{Krishak},
  {Dantuluri}, and {Desai}}}]{Krishak}
\bibinfo{author}{\bibfnamefont{A.}~\bibnamefont{{Krishak}}},
  \bibinfo{author}{\bibfnamefont{A.}~\bibnamefont{{Dantuluri}}},
  \bibnamefont{and} \bibinfo{author}{\bibfnamefont{S.}~\bibnamefont{{Desai}}},
  \bibinfo{journal}{\jcap} \textbf{\bibinfo{volume}{2020}}, \bibinfo{eid}{007}
  (\bibinfo{year}{2020}), \eprint{1906.05726}.

\bibitem[{\citenamefont{{Krishak} and {Desai}}(2019)}]{Krishak2}
\bibinfo{author}{\bibfnamefont{A.}~\bibnamefont{{Krishak}}} \bibnamefont{and}
  \bibinfo{author}{\bibfnamefont{S.}~\bibnamefont{{Desai}}},
  \bibinfo{journal}{The Open Journal of Astrophysics}
  \textbf{\bibinfo{volume}{2}}, \bibinfo{pages}{E12} (\bibinfo{year}{2019}),
  \eprint{1907.07199}.

\bibitem[{\citenamefont{Adhikari et~al.}(2019)}]{Cosine}
\bibinfo{author}{\bibfnamefont{G.}~\bibnamefont{Adhikari}} \bibnamefont{et~al.}
  (\bibinfo{collaboration}{COSINE-100}), \bibinfo{journal}{Phys. Rev. Lett.}
  \textbf{\bibinfo{volume}{123}}, \bibinfo{pages}{031302}
  (\bibinfo{year}{2019}), \eprint{1903.10098}.

\bibitem[{\citenamefont{{Sharma}}(2017)}]{Sanjib}
\bibinfo{author}{\bibfnamefont{S.}~\bibnamefont{{Sharma}}},
  \bibinfo{journal}{\araa} \textbf{\bibinfo{volume}{55}}, \bibinfo{pages}{213}
  (\bibinfo{year}{2017}), \eprint{1706.01629}.

\bibitem[{\citenamefont{{Liddle}}(2004)}]{Liddle}
\bibinfo{author}{\bibfnamefont{A.~R.} \bibnamefont{{Liddle}}},
  \bibinfo{journal}{\mnras} \textbf{\bibinfo{volume}{351}},
  \bibinfo{pages}{L49} (\bibinfo{year}{2004}), \eprint{astro-ph/0401198}.

\bibitem[{\citenamefont{{Liddle}}(2007)}]{Liddle07}
\bibinfo{author}{\bibfnamefont{A.~R.} \bibnamefont{{Liddle}}},
  \bibinfo{journal}{\mnras} \textbf{\bibinfo{volume}{377}},
  \bibinfo{pages}{L74} (\bibinfo{year}{2007}), \eprint{astro-ph/0701113}.

\bibitem[{\citenamefont{{Shi} et~al.}(2012)\citenamefont{{Shi}, {Huang}, and
  {Lu}}}]{Shi}
\bibinfo{author}{\bibfnamefont{K.}~\bibnamefont{{Shi}}},
  \bibinfo{author}{\bibfnamefont{Y.~F.} \bibnamefont{{Huang}}},
  \bibnamefont{and} \bibinfo{author}{\bibfnamefont{T.}~\bibnamefont{{Lu}}},
  \bibinfo{journal}{\mnras} \textbf{\bibinfo{volume}{426}},
  \bibinfo{pages}{2452} (\bibinfo{year}{2012}), \eprint{1207.5875}.

\bibitem[{\citenamefont{{Trotta}}(2017)}]{Trotta}
\bibinfo{author}{\bibfnamefont{R.}~\bibnamefont{{Trotta}}},
  \bibinfo{journal}{arXiv e-prints} \bibinfo{eid}{arXiv:1701.01467}
  (\bibinfo{year}{2017}), \eprint{1701.01467}.

\bibitem[{\citenamefont{{Ivezi{\'c}} et~al.}(2014)\citenamefont{{Ivezi{\'c}},
  {Connolly}, {Vanderplas}, and {Gray}}}]{astroml}
\bibinfo{author}{\bibfnamefont{{\v Z}.}~\bibnamefont{{Ivezi{\'c}}}},
  \bibinfo{author}{\bibfnamefont{A.}~\bibnamefont{{Connolly}}},
  \bibinfo{author}{\bibfnamefont{J.}~\bibnamefont{{Vanderplas}}},
  \bibnamefont{and} \bibinfo{author}{\bibfnamefont{A.}~\bibnamefont{{Gray}}},
  \emph{\bibinfo{title}{Statistics, Data Mining and Machine Learning in
  Astronomy}} (\bibinfo{publisher}{Princeton University Press},
  \bibinfo{year}{2014}).

\bibitem[{\citenamefont{{Kerscher} and {Weller}}(2019)}]{Weller}
\bibinfo{author}{\bibfnamefont{M.}~\bibnamefont{{Kerscher}}} \bibnamefont{and}
  \bibinfo{author}{\bibfnamefont{J.}~\bibnamefont{{Weller}}},
  \bibinfo{journal}{SciPost Physics Lecture Notes} \textbf{\bibinfo{volume}{9}}
  (\bibinfo{year}{2019}), \eprint{1901.07726}.

\bibitem[{\citenamefont{{Messina} et~al.}(2020)\citenamefont{{Messina},
  {Nardecchia}, and {Piacentini}}}]{Messina}
\bibinfo{author}{\bibfnamefont{A.}~\bibnamefont{{Messina}}},
  \bibinfo{author}{\bibfnamefont{M.}~\bibnamefont{{Nardecchia}}},
  \bibnamefont{and}
  \bibinfo{author}{\bibfnamefont{S.}~\bibnamefont{{Piacentini}}},
  \bibinfo{journal}{\jcap} \textbf{\bibinfo{volume}{2020}}, \bibinfo{eid}{037}
  (\bibinfo{year}{2020}), \eprint{2003.03340}.

\bibitem[{\citenamefont{{Ganguly} and {Desai}}(2017)}]{Ganguly}
\bibinfo{author}{\bibfnamefont{S.}~\bibnamefont{{Ganguly}}} \bibnamefont{and}
  \bibinfo{author}{\bibfnamefont{S.}~\bibnamefont{{Desai}}},
  \bibinfo{journal}{Astroparticle Physics} \textbf{\bibinfo{volume}{94}},
  \bibinfo{pages}{17} (\bibinfo{year}{2017}), \eprint{1706.01202}.

\bibitem[{\citenamefont{Wilks}(1938)}]{Wilks}
\bibinfo{author}{\bibfnamefont{S.~S.} \bibnamefont{Wilks}},
  \bibinfo{journal}{Annals Math. Statist.} \textbf{\bibinfo{volume}{9}},
  \bibinfo{pages}{60} (\bibinfo{year}{1938}).

\bibitem[{\citenamefont{{Lyons}}(2016)}]{Lyons}
\bibinfo{author}{\bibfnamefont{L.}~\bibnamefont{{Lyons}}},
  \bibinfo{journal}{arXiv e-prints} \bibinfo{eid}{arXiv:1607.03549}
  (\bibinfo{year}{2016}), \eprint{1607.03549}.

\bibitem[{\citenamefont{{Protassov} et~al.}(2002)\citenamefont{{Protassov},
  {van Dyk}, {Connors}, {Kashyap}, and {Siemiginowska}}}]{Kashyap}
\bibinfo{author}{\bibfnamefont{R.}~\bibnamefont{{Protassov}}},
  \bibinfo{author}{\bibfnamefont{D.~A.} \bibnamefont{{van Dyk}}},
  \bibinfo{author}{\bibfnamefont{A.}~\bibnamefont{{Connors}}},
  \bibinfo{author}{\bibfnamefont{V.~L.} \bibnamefont{{Kashyap}}},
  \bibnamefont{and}
  \bibinfo{author}{\bibfnamefont{A.}~\bibnamefont{{Siemiginowska}}},
  \bibinfo{journal}{\apj} \textbf{\bibinfo{volume}{571}}, \bibinfo{pages}{545}
  (\bibinfo{year}{2002}), \eprint{astro-ph/0201547}.

\bibitem[{\citenamefont{{Cowan} et~al.}(2011)\citenamefont{{Cowan}, {Cranmer},
  {Gross}, and {Vitells}}}]{Cowan}
\bibinfo{author}{\bibfnamefont{G.}~\bibnamefont{{Cowan}}},
  \bibinfo{author}{\bibfnamefont{K.}~\bibnamefont{{Cranmer}}},
  \bibinfo{author}{\bibfnamefont{E.}~\bibnamefont{{Gross}}}, \bibnamefont{and}
  \bibinfo{author}{\bibfnamefont{O.}~\bibnamefont{{Vitells}}},
  \bibinfo{journal}{European Physical Journal C} \textbf{\bibinfo{volume}{71}},
  \bibinfo{eid}{1554} (\bibinfo{year}{2011}), \eprint{1007.1727}.

\bibitem[{\citenamefont{{Speagle}}(2020)}]{Speagle}
\bibinfo{author}{\bibfnamefont{J.~S.} \bibnamefont{{Speagle}}},
  \bibinfo{journal}{\mnras}  (\bibinfo{year}{2020}), \eprint{1904.02180}.

\end{thebibliography}
\end{document}